\documentclass[nofootinbib,superscriptaddress,twocolumn]{revtex4}

\begin{document}

\pagestyle{plain}

\title{Opportunities in Neutrino Theory -- a Snowmass White Paper}

\author{Andr\'e de Gouv\^ea\footnote{email: degouvea@northwestern.edu}}
\affiliation{Department of Physics \& Astronomy, Northwestern University, IL 60208-3112, USA}
\author{Alexander Friedland\footnote{email: friedland@lanl.gov}}
\affiliation{Theoretical Division T-2, MS B285, Los Alamos National Laboratory, Los Alamos, NM 87545, USA}
\author{Patrick Huber\footnote{email: pahuber@vt.edu}}
\affiliation{Center for Neutrino Physics, 
  Virginia Tech, Blacksburg, VA 24061, USA}
\author{Irina Mocioiu\footnote{email: irina@phys.psu.edu}}
\affiliation{Department of Physics, The Pennsylvania State University, University Park, PA 16802, USA}


\begin{abstract}
Neutrino masses are clear evidence for physics beyond the standard
model and much more remains to be understood about the neutrino
sector.  We highlight some of the outstanding questions and research
opportunities in neutrino theory. We show that most of these questions
are directly connected to the very rich experimental program currently
being pursued (or at least under serious consideration) in the United
States and worldwide. Finally, we also comment on the state of
the theoretical neutrino physics community in the U.S.
\end{abstract}
\maketitle

\section{Introduction}

No area of fundamental particle physics research changed more
dramatically or rapidly over the past two decades than neutrino
physics.\footnote{The nature of this document is such that no
  references to the literature will be included. }
It was demonstrated, in a spectacular and conclusive fashion, that the
long-standing solar and atmospheric neutrino ``problems'' were in fact
caused by neutrino flavor oscillations. Moreover, the mass splittings
and the three-flavor mixing parameters have been since established
with impressive accuracy and the subject has entered a high-precision
era.

These discoveries, in turn, opened the door to a rich, exciting, and
promising research program aimed at elucidating the origin of neutrino
masses, exploring the flavor structure of the lepton sector, pursuing
new sources of CP-invariance violation, looking for new particles and
interactions, and testing fundamental physics principles. Nonzero
neutrino masses also modify, sometimes dramatically, the role of
neutrinos in cosmology and astrophysics.

Experimentally, the path forward is well defined: a diverse neutrino
program is required in order to explore the new physics revealed in
the neutrino sector. It includes, necessarily, very intense neutrino
beams, very large (tens of kilotons and above), finely instrumented
detectors with different detection media (water, argon, hydrocarbons,
etc), very large, ultra clean detectors to search for neutrinoless
double-beta decay, novel detectors for precision measurements of
beta-decay, etc. In the next decades, a deluge of neutrino-related
data is expected. Neutrino theorists and phenomenologists will be in
the enviable position of exploiting these unique probes of fundamental
physics, interpreting the data, building models to accommodate new
phenomena and connecting the new discoveries in neutrino physics to
other areas of particle physics, astrophysics, and cosmology.

In the next section, we briefly discuss a sample subset of the many
outstanding opportunities in `neutrino theory,' most of which are
either directly or indirectly related to, or facilitated by, the
discovery of nonzero neutrino masses and ongoing efforts to perform
precision measurements of the neutrino sector. The breadth of topics
is noteworthy, ranging from collider physics to cosmology and
astrophysics, from nuclear physics to grand unification. Most of the
activities are synergistic with the current and future neutrino
experimental programs, and many require specialized technical and
theoretical skills that are not encountered in abundance within the
theoretical particle physics community in the United States. We
comment on this last point more concretely in the last section.

\section{Neutrino Theory: Select Outstanding Questions and Opportunities}

In order to illustrate the richness and diversity of scientific
endeavors in neutrino physics and to highlight the close connection to
the planned experimental program, we briefly discuss a sample of
select physics topics.

\subsection{Understanding the Origin of Neutrino Masses -- Model Building and Phenomenology}

Neutrino masses represent one of the very few experimental clues
regarding the physics that lies beyond the standard model. Given the
standard model gauge symmetries and particle content, new degrees of
freedom are necessarily required, and the symmetry structure of the
Lagrangian must be modified qualitatively. The dynamics behind
neutrino masses is currently unknown. The number of options is large
and incredibly diverse, and so are the many distinct, potentially
observable, signatures associated with the different options. As an
example, neutrino masses may be a consequence of physics that violates
lepton number, either explicitly or spontaneously. The current data
allow virtually any value for the lepton-number breaking scale, from a
few eV to $10^{15}$~GeV -- twenty-four orders of magnitude.

Extensive theory effort is required in order to identify the different
models that lead to non-zero neutrino masses and to work out the
relevant phenomenology, both within and outside of neutrino
physics. The mechanism behind neutrino masses, it is well-known, can
manifest itself in a variety of different observables including
lepton-number violating processes like neutrinoless double-beta decay,
forbidden meson decays (e.g. $K^+\to\pi^-\mu^+\mu^+$), and high energy
collisions (e.g., $pp\to{\rm jets}~e^+\mu^+$ and no missing energy),
lepton-flavor number violating processes like $\mu\to e$-conversion in
nuclei, $\tau\to\mu\mu\mu$, and $K_L\to e \mu$. A sufficiently low
scale of neutrino mass generation may result in measurable
``nonstandard'' neutrino interactions, which may reveal themselves in
a variety of experimental setups, from neutrino oscillations to LHC
collisions, as described below. Additionally, the mechanism
responsible for neutrino masses often predicts additional, ``sterile''
neutrino states, which could be revealed in precision oscillation
experiments.

More broadly, the mechanism behind neutrino masses may also be related
to other fundamental problems in particle physics, including the
nature and origin of the dark matter, the dynamics responsible for the
matter--antimatter asymmetry of the universe, electroweak symmetry
breaking, and grand unification. Theory work is required in order to
identify potential connections and explore all corroborating
phenomenological consequences. Well-known examples include the
possibility that nonzero Majorana neutrino masses are a consequence of
a more complicated Higgs sector (e.g., the so-caled Type-II seesaw),
the hypothesis that the dark matter is related to right-handed
neutrinos (which can be probed, for example, by looking for cosmic
X-rays from dark matter decays), and baryogenesis via leptogenesis, a
natural consequence of the so-called Type-I seesaw mechanism, as long
as the lepton-number breaking scale is high
enough. Phenomenologically, leptogenesis in particular provides a most
interesting challenge: under what conditions can it be falsified (or
``confirmed'')?

\subsection{Flavor Models, CP Violation in the Lepton Sector} 

In the quark sector, the pattern of masses and mixing parameters seems
to hint at the existence of some yet-to-be-uncovered organizing
principle. Such an organizing principle remains unknown, in spite of
half of century of theoretical work. The discovery of nonzero neutrino
masses and lepton mixing added new pieces to the flavor puzzle. In
particular, the leptonic mixing matrix appears to be providing
qualitatively different information. Unlike the quark mixing matrix,
it cannot be understood as an identity matrix ÒperturbedÓ by small,
hierarchical, off-diagonal elements.

Precision neutrino oscillation experiments provide a unique
opportunity to test different ideas in flavor physics. Next-generation
data will rule out a large subset of the parameter space of flavor
models, and will guide the next-generation of flavor physics. On the
flip side, flavor models are necessary in order to provide guidance
regarding precision milestones for neutrino oscillation experiments. A
concrete recent example: the discovery of a ``large'' value for
$\theta_{13}$ ``ruled out'' many flavor paradigms, and triggered
interest in the precise value of $\theta_{23}$, especially when it
comes to its deviation from maximal. Identifying whether $\theta_{23}$
deviates from maximal ``at the $\theta_{13}$ level'' is also a very
stringent test of several classes of flavor models. In addition,
flavor physics allows one to relate, with the help of ``more'' new
physics, different types of observables, including measurements of
oscillation parameters, the rates of different charged-lepton flavor
violating processes, and the masses and mixing patterns of new
particles that may manifest themselves at the LHC or future
high-energy collider enterprises.

Developments in neutrino physics provide another unique opportunity
for theoretical physics: a new CP-violating sector. With nonzero
neutrino masses, the $\nu$ Standard Model Lagrangian -- whatever it is
-- accommodates, if the neutrinos are Dirac (Majorana) fermions, at
least three (four) CP-violating parameters. They all appear to be
unrelated, and only two are known. The CKM phase $\delta$ is around
$\pi/2$ (hence quite large), while the strong CP ``phase'' is either
zero or less than $10^{-9}$ (hence tiny). The study of CP-invariance
violating phenomena in the lepton sector, which is possible in precise
long-baseline neutrino oscillation experiments,\footnote{More
  CP-violating parameters can be studied via neutrino oscillations if
  there are more neutrino states (sterile neutrinos). If neutrinos are
  Majorana fermions, the so-called Majorana phases are also physical
  but very hard to study experimentally, as they usually manifest
  themselves only in observables that violate lepton number.} will
open a new window on CP-violation and may shed light on its potential
origin.

\subsection{Non-standard Interactions, Neutrino Decays, Neutrino Electromagnetic Properties}

As mentioned earlier, the existence of neutrino masses point towards
new physics beyond the Standard Model, and perhaps one or more new
mass scales.
The values of such new scales are presently unknown. It is of great importance to understand whether there are other observable effects associated to these scales.

A general strategy is to search for other operators that could be
generated at the new scale(s). Such operators are expected on very
general grounds, but whether their effects are observable depends on
how large is the new energy scale. Their effects range from
unobservably tiny, if the corresponding new physics scale is very
high, say, $10^{14}$ GeV, to within experimental reach, if the new
scale is near a TeV or below. Given all presently available data, the
latter possibility is perfectly allowed. The potentially revolutionary
impact the discovery of new neutrino interactions renders the
exploration of this possibility is mandatory.

Potentially observable effects include, in addition to some the
processes mentioned in Sec. IIA, new ``four-fermion'' neutrino
interactions with quarks and leptons (NSI), or electromagnetic
interactions of the neutrino. NSI modify neutrino oscillation
probabilities in matter via their contributions to the so-called
matter potential, and potentially affect the production and detection
processes. NSI may also manifest themselves in non-oscillation
experiments, including the high energy colliders (in, for example, the
monojet searches).

There are many issues associated with NSI still to be understood, and
many possible approaches to doing so. One approach is to start from
the model building side and try to understand all possible
implications of a given model, as discussed above.  Another
possibility is to adopt the phenomenological approach where the
effects of new interactions are parametrized by a general set of free
parameters that can be constrained by data.  In the model building
context it is important to correlate predictions for neutrino
oscillation parameters with the implications of the model in other
directions like charged lepton measurements, and astrophysical or
cosmological observations.

When adopting the phenomenological approach, one way to identify NSI
is to look for inconsistencies of the three-flavor paradigm once one
compares data from different experiments. This requires the ability to
carefully and consistently combine different data sets, as will
discussed more fully in the next section.  Even if discrepancy are not
observed, the presence of new interactions can affect the
reconstruction of standard oscillation parameters. We understand, for
example, that the NSI can introduce large degeneracies into the
problem. A detailed quantitative analysis of such issues, however, is
still missing.

Massive neutrinos couple to the electromagnetic field at the quantum
level, and such couplings may lead to potentially observable effects
in neutrino physics and astrophysics. Neutrino magnetic (transition)
moments affect, for example, the flavor evolution of solar neutrinos
in the solar magnetic field.
literature.  For many years, the interest in this possibility was
driven by the hint of modulation of the Homestake event rate with the
solar cycle. Nowadays, with precision solar neutrino data at hand, it
is possible to ask whether such effects exist at the subdominant
levels.  It is also of great interest to understand what effect it may
have on supernova neutrinos. The electromagnetic moments can also have
an impact on the evolution of stars, as they can increase the rate of
energy loss from stellar cores. In fact, the most stringent known
bounds on the neutrino magnetic models come from the observations and
modeling of red giant stars before helium flash.

Successful research efforts require a broad view of neutrino, particle
physics, astrophysics and cosmology, making the connections between
specific models trying to understand fermion mass origin and flavor
physics, global fits to neutrino data and possible new implications of
any new physics.

\subsection{Neutrino Oscillation Phenomenology -- Measuring Neutrino Properties and Looking for New Physics}

Neutrino physics has been a data-driven field for most of its
existence and phenomenology has always played a central role in
advancing our understanding. Confronting the prevailing theoretical
paradigm with data in the case of neutrinos has often led to dramatic
revisions of the theoretical world view. What evolved into
overwhelming evidence for neutrino oscillations started out as the
``solar neutrino problem'', which was, at first, attributed to
everything but the underlying neutrino properties. The belief that
neutrinos were massless or that their mixing angles were small were
crushed by experimental data in a rather dramatic fashion. In all of
these cases, combined analyses of prior data -- ``global fits'' --
guided the design of definitive experiments, such as SNO and KamLAND.

Presently, there are numerous experiments probing neutrino
oscillations at very different energies and baselines, including the
reactor experiments Daya Bay, RENO, and Double-CHOOZ and the beam
neutrino experiments T2K, MINOS(+), and NO$\nu$A. To fully understand
the implications of the measurements from these experiments, they need
to be combined with one another, as well as with the solar neutrino
data from SNO, SuperKamiokande, Borexino and GALLEX/SAGE, the
atmospheric neutrino data from SuperKamiokande, the short-baseline
data from LSND, Mini-BOONE, etc. The important point is that no single
experiment dominates the determination of the entire oscillation
matrix.

To further illustrate the value of global fits, consider the situation
with the so-called ``short-baseline'' anomalies. Currently, we are
confronted with a set of anomalies from a rather diverse set of
experiments, including beams and reactors. Each single anomaly is at
about the $3\,\sigma$ confidence level and, interpreted in isolation,
does not amount to much. In combination, however, they may be pointing
to the same region of the parameter space, if one interprets them as
evidence for one or more sterile neutrinos.  While more experimental
and theoretical work is required in order to resolve these
short-baseline anomalies, it is clear that combined analyses of data,
performed outside of any experimental collaboration, have provided
very valuable information.

Going forward, the need for global fits is likely to persist. They are
expected to play a central role in delivering answers to the most
central questions of the next decade and likely beyond. CP-violation
searches, for example, will rely, in the foreseeable future, on
interpreting he combination of data from T2K, NO$\nu$A, LBNE, and
other experiments. Moreover, combined fit analyses will be essential
to search for deviations from the basic three-flavor paradigm which
may arise, for example, from nonstandard interactions, as mentioned
earlier. Global fits to all neutrino data will also play a fundamental
role in addressing questions like the unitarity of the neutrino mixing
matrix and testing predictions from a variety of flavor models.

Worldwide, there is a handful of groups dedicated to performing
high-quality global fits. In the US there are no research groups
performing this very technical (among the necessary skill are
statistical analysis of very different data sets and a detailed
understanding of neutrino flavor oscillations within and outside the
standard three-flavor paradigm), but very impactful task.  With the US
poised to become the world-leader in long-baseline oscillation
experimentation, it is highly desirable to build up this capability,
to ensure we are able to extract as much information from the data as
possible and to identify future directions for neutrino oscillation
research.

\subsection{Accurate Computation and Parameterization of Neutrino Nucleus Cross Sections}

Despite the fact that the electroweak sector of the Standard Model is
extremely well understood, it is very hard to perform precise
computations of the neutrino-nucleus interactions. The nucleons are
composite states with complex dynamics, which so far has eluded first
principles calculations, despite QCD being a full description of the
underlying physics. To make matters worse, neutrino detectors are not
made of free nucleons but of materials consisting of nuclei covering a
large atomic mass range from $A=12$ carbon (scintillator) over $A=40$
argon (TPCs) to $A=56$ iron (calorimeters). Since nuclear structure is
not well understood, especially in large nuclei, it is presently not
possible to formulate a closed theory of neutrino-nucleus
interactions. Moreover, multi-nucleon correlations as well as final
state interactions have to be correctly included to provide reliable
neutrino cross sections.
 
 The situation is, however, not hopeless.  It is possible, for
 example, to derive the correct initial state densities by exploiting
 the symmetries of the electroweak theory and using existing data from
 electron scattering -- the results known as the spectral
 functions. However, to this date there is no complete and validated
 implementation of spectral functions in any publicly accessible event
 generator. More complicated issues, like meson exchange currents are
 even lacking a consensus theory formulation and are years away from
 being available in event generators.

 At the same time, the next generation of neutrino oscillation
 experiments aim at percent level measurements of neutrino event rates
 and precise measurements of the energy distribution of events. Since
 existing neutrino beams are subject to large intrinsic uncertainties,
 neutrino--nucleus cross sections are only known at the the 10-30\%
 level. Many nuclear effects have non-trivial energy dependencies and
 introduce large biases in neutrino energy reconstruction and
 therefore, most experiments really measure an effective cross section
 specific to the beam and energy response of the detector. This is
 also the reason that near detectors will not resolve all, or even the
 majority of the issues. At the time of this writing the largest
 contribution to the systematic error budget of T2K comes from
 neutrino interactions. The situation in the upcoming NO$\nu$A and
 LBNE experiments is expected to be at least as challenging.

Contributions to this effort require mastery of both precision nuclear
and particle physics calculations, and in-depth understanding of
detectors and event generator codes.  The current theory effort, in
the US, dedicated to attacking the challenges described above (and
many others) is close to non-existent and starkly different from the
large investment in experiments like NO$\nu$A and LBNE.  In absence of
an adequate theory effort dedicated to the understanding of neutrino
scattering on nuclei, NO$\nu$A and LBNE will not be able to provide
the world-class science they were designed for.

\subsection{Supernova Neutrinos}

The utility of core-collapse supernovae for particle physics is well recognized. For example, upon perusing the Particle Data Group summaries of various Beyond-the-Standard-Model (BSM) searches, one encounters numerous constraints on new physics scenarios derived from supernova cooling considerations. 
The cooling argument, in brief, is as follows. To account for the observed duration of the neutrino pulse from the 1987A supernova, the energy trapped in the collapsed core needs to get out on the time scale of a few seconds. It can be easily seen that ``normal" astrophysical methods of transport, such as photon diffusion, are too slow for this: the photon gets stuck in the dense matter of the core because of its large scattering cross section. Neutrinos, particles with \emph{much smaller} cross sections, take up the transport task. Indeed, a straightforward estimate for neutrinos of $10^7$ eV energies suggests a diffusion time scale of a few seconds.  It is instructive that neutrinos ``win'' over photons by having \emph{smaller} cross sections, a situation that is highly counterintuitive by laboratory standards. Any putative new particle with a still smaller cross section would be even more efficient at carrying the energy out, so long as such a particle couples strongly enough to be produced in the core at all.
Examples of such bounds include well-known constraints on axion-nucleon coupling, Majorons, Kaluza-Klein gravitons, unparticles, and extra-dimensional photons.

It should be stressed that the present constraints are based on the 1987 supernova observations, which gave us only two dozen neutrino events. Given this very limited statistics, rough order-of-magnitude statements about what new physics can be excluded are, for the most part, sufficient. For other scenarios, rough estimates are woefully inadequate. This is particularly true when the new physics ``only'' modifies neutrino transport by order one factors, rather than completely taking over. 
Furthermore, the next event should yield many more neutrinos. This is so, firstly, because the next supernova will, much more likely, take place in our Galaxy rather than in the Large Magellanic Could (and an $r^2$ enhancement factor of $\sim$ 10-100 is to be reasonably expected), and secondly, because we now have much bigger detectors than what was available in 1987 (Kamiokande and IMB). 

With such abundant data, to properly interpret the neutrino signal, one will need to take into account neutrino flavor oscillations. Rather than being simply a nuisance, neutrino oscillations in a supernova environment represent an extremely rich source of physical information. The subject of supernova neutrino oscillations has undergone dramatic progress in the last ten years and more effects continue to be uncovered every year. For example, these oscillations can be impacted by the developing explosion: the expanding shock front and the turbulent region behind it change the density profile in which neutrinos change flavor. As the character of the transformation is changed (for example, from adiabatic to non-adiabatic by the front shock, or to an incoherent sum of states by the turbulent density fluctuations), an imprint should be left on the neutrino signal. Thus, properly reading the flavor-transformed signal can tell us how the explosion develops. 

The uniqueness of the supernova environment is perfectly illustrated by the fact that it can host neutrino ``self-induced'' transformations, which are hopelessly inaccessible in the laboratory. Also known as ``collective oscillations'', these transformations happen when the density of streaming neutrinos is so high that their flavor evolutions become coupled. In the last decade, the available computing power made it possible to explore this novel many-body phenomenon and remarkable flavor-transformation patterns were uncovered. An observational confirmation of this dynamics would be a discovery of profound magnitude. 

Taking a broader view, we must remember that supernova neutrinos are of great interest to many areas of physics. For example, in nuclear astrophysics, the fundamental question is the origin of heavy elements in the universe. Core-collapse supernovae are thought to be candidate sites for the $r$-process and other types of nucleosynthesis. The efficiency of the $r$-process depends on the physical conditions in the explosion, such as the entropy profile of the neutrino-driven wind and the energy spectra of the different neutrino components (which in turn depend on the oscillation physics). It is of great interest whether the future neutrino signals can shine light on the nucleosynthesis mechanism in core-collapse supernovae.

Supernovae are also of fundamental importance in astrophysics and cosmology, where they controll baryonic structure formation and evolution. They do so by creating and spreading heavy elements (``metals''), seeding star formation by shocks, blowing out gas from small gravitational potentials, etc. For all these reasons, the studies of the explosion mechanism have been occupying astrophysicists for over half a century. 

The focal question for neutrino theorists here is to understand how different explosion mechanisms and new physics effects can manifest themselves in the neutrino signal and what detector characteristics are required to measure them.
Specifically, one needs to identify what key, ``smoking-gun'' signatures to look for and what detector characteristics are necessary. For example, to test for the presence of new particles or new neutrino interactions, it is important to measure the neutrino energy spectra and to track their time evolution. To observe collective oscillations, one needs detailed flavor information, in both neutrino and antineutrino channels. Simply bigger detectors would thus not be enough. It is also important to have sensitivity to different neutrino flavors, as well as good energy resolution. While it is not known when the next galactic supernova will be observed, it would be a travesty to be caught unprepared for this ``once-in-a-lifetime'' opportunity.

To tackle these and several similar problems, a generation of theorists with broad education in particle physics, nuclear physics, nucleosynthesis, plasma physics, turbulence, transport, supercomputing, etc, is required. It is also imperative to develop  a common language between the neutrino theorists and the experimentalists designing and simulating detectors. Clearly, as a community, we have our work cut out for us.

\subsection{Neutrinos and Cosmology}

Neutrinos are also well known to play a key role in cosmology, through their impact on the evolution of cosmic perturbations and growth of structure. Early on, during the radiation dominated epoch, the cosmic neutrino background was an important component of the gravitating matter in the universe. They, therefore, had a direct impact on the expansion rate before and during the cosmic microwave background (CMB) decoupling era. Additionally, neutrinos affected the evolution of density perturbations. In this role, neutrinos were quite different from photons: the latter stayed coupled to the plasma, while the former streamed out of the perturbations once allowed by causality. 

All of this physics has recently become experimentally accessible, thanks to the data from the Planck satellite and terrestrial observatories such as ACT, SPT, etc. The Planck satellite, in particular, has presented constraints on $N_{\rm eff}$  -- the number of neutrino-like relativistic species in the early universe. The constrains are extremely relevant for models which extend the neutrino sectors to include new states ({\it e.g.}, ``sterile neutrinos''). It should be noted, however, that there are a lot of open theoretical questions. For example, what if cosmology is less trivial ({\it i.e.}, goes beyond the ``vanilla'' 6-parameter study)? What if the extra neutrino-like species have non-thermal spectra? What if their abundance in the universe changes with time? As the measurements of $N_{\rm eff}$ enter a precision stage, these are the right questions to ask. For example, there are models of neutrino ``recoupling'', in which the number of new neutrino species appearing in CMB and big-bang nucleosynthesis calculations are different. 

Neutrinos also show up in precision cosmological observations in another way: since they have a small mass, they should cluster on sufficiently large scales. This phenomenon should result in potentially observable CMB \emph{lensing} signatures. It is expected that these signatures could allow us to probe very small neutrino masses, possibly bellow the  0.1~eV level suggested by the atmospheric oscillation data. Again, there are numerous theoretical questions to study in connection with this, for example, how these measurements would be sensitive to nonstandard physics in the neutrino sector. 

Recently, a number of provocative ideas have been put forth on additional possible roles neutrinos could play in cosmology. As an illustration, if neutrinos couple to dark matter, they could alter the formation of structure, potentially alleviating the ``missing satellite'', or ``too big to fail'' problems. Neutrinos may also be messengers of dark matter annihilation, in our galactic halo, or in the core of the Sun. 

With precision cosmological observations and oscillation experiments planned for the next 10-20 years, numerous potentially interesting developments may become possible. For example, suppose the next-generation oscillation experiments confirm the existence of sterile neutrinos, in a way that is seemingly incompatible with cosmologically inferred value for $N_{\rm eff}$. This conflict would be an indication of nonstandard cosmology and/or new physics in the neutrino sector and the task will be to understand how to best disentangle this situation. It is prudent to anticipate and prepare for such nonstandard scenarios now, as the experiments are being planned. Work in this field requires deep understanding of various aspects of cosmological physics (CMB, LSS, etc), laboratory neutrino data, as well as particle physics broadly defined.

\subsection{Phenomenology of Astrophysical High-Energy and Ultra-High-Energy Neutrinos} 

Very high energy neutrinos are predicted to be produced by a variety of astrophysical sources and neutrino telescopes are actively looking  
for them. This year, the IceCube detector has brought the first observational indication of such astrophysical neutrinos. The collaboration has shown  
evidence for two events at PeV energies, as well as 28 events at slightly lower energy, a $4 \sigma$ excess above atmospheric neutrino  
background. In the coming year, an additional  data sample will be analyzed and the detector will continue accumulating new data. These exciting developments make theoretical  
study of the particle physics and astrophysics of the very high energy neutrinos quite urgent.

The physics and astrophysics that can be probed by high energy neutrinos is extremely rich. The relative effects of new physics on such signals can be large: thanks to very long propagation distances, 
even very small new physics effects can potentially add up  
to an observable signal. Moreover, the extremely high energies can provide access to the energy regimes that cannot be reached in colliders  or other terrestrial experiments. On the astrophysics front, the weak interactions allow neutrinos to probe environments that are not otherwise accessible and may help answer many long-standing questions about the origin, composition and propagation of cosmic rays.


Observation of high energy neutrinos will likely have profound implications for both physics and astrophysics. Neutrino telescopes looking for such neutrinos already exist or are being planned to cover even higher energy ranges. The theoretical studies necessary to  
extract all the information from the observations and understand the signals and the various uncertainties have just begun. Much work is still  
to be done, including better modeling of sources, propagation and detection, considering all possible standard physics, astrophysics and  
new physics possibilities, as well as new detection techniques. Additional questions are likely to arise in the course of these analyses. All this work will require 
a good understanding of astrophysics, modeling of strong interactions (relevant for the production and detection of the neutrinos), neutrino 
physics and possible new physics, as well as a good understanding of what these telescopes can actually measure and correlations between different types of observables in neutrino, cosmic ray, astrophysical as well as other particle physics measurements.

\section{Neutrino Theory in the US: Where We Stand and Paths Forward}

While the opportunities in neutrino theory are many and the range of topics broad and interdisciplinary, the fraction of the US domestic theoretical physics effort dedicated to neutrinos is small. Currently, only very few university groups are active in neutrino phenomenology. Indeed, since 1998  -- the year neutrino flavor change was unambiguously discovered by Super-Kamiokande -- less than 10\% of all particle theory university hires\footnote{This was extracted from the ``Theoretical Particle Physics Jobs Rumor Mill,'' \url{http://particle.physics.ucdavis.edu/rumor/doku.php}, and private communications with members of the theory community.} were on neutrino theory or broadly defined related areas. 

At the same time, with the start of the LHC in 2010 and the shutdown of the Tevatron in 2011, the focus of the US domestic experimental particle physics program has shifted to the so-called Intensity Frontier. Neutrino physics makes up a big fraction of the Intensity Frontier research currently being discussed. In order to ensure a vital program, the funding agencies and the community appear committed to making very significant investments into new experiments at the Intensity Frontier over the next decade. As we have illustrated in this white paper, this experimental program will require a commensurate level of activity on the theory side to ensure that this investment results in a corresponding return in science. This work ranges from identifying physics goals and desired detector characteristics to accurate theoretical calculations of expected signals and backgrounds, without which the experiments cannot succeed. This means that a much stronger US presence in neutrino theory in particular and Intensity Frontier phenomenology in general is not only scientifically well-motivated but also necessary to ensure the long-term success of the US  domestic particle physics program. 

The combination of a currently small neutrino theory community and the fact that neutrino theory research requires technical and theoretical skills -- as exemplified above -- outside the standard ``tool-kit'' of the US particle theory community implies that the neutrino theory community will not grow ``organically'' from within the current  theory community on the short time scale required by the planned experimental program. A dedicated and coherent effort, with (a) significant investment from the funding agencies, (b) enthusiastic commitment and leadership from the present neutrino theory community, and (c) the support of the entire particle theory community is absolutely necessary. To meet the proposed experimental schedules, such a qualitative increase in the neutrino theory effort is needed now. 


An informal `Neutrino Theory' meeting took place on May 20, 2013 at Fermilab in order to foster discussion regarding the status of theoretical neutrino physics in the US and potential initiatives to strengthen and increase the domestic neutrino and Intensity Frontier theoretical communities. A little under twenty theorists attended the meeting while many other members of the theory community provided feedback and support via email. The issues presented in this White Paper are partially informed by these discussions and feedback collected during the last several months. It is anticipated that more discussions will follow during the next several months. Ultimately, the goal is to converge on a concrete ``project'' aimed at qualitatively strengthening the US ``neutrino theory'' community in order to take full advantage of the opportunities described here, along with many more that are sure to present themselves. 
\\

\begin{acknowledgments}
We are indebted to all of those that participated, directly or indirectly, in the several ``neutrino theory'' and related discussions that took place during the last several months, including the Neutrino Working Group Meeting at SLAC (March 2013), the Intensity Frontier Workshop at ANL (April 2013), the Neutrino Theory Meeting at FNAL (May 2013), and the Community Planning Meeting in Minneapolis (August 2013). 
\end{acknowledgments}

\bibliographystyle{apsrev} 
\bibliography{references}
\end{document}